\documentclass[]{tMOP2e}
\usepackage{color}

\begin{document}

\doi{10.1080/0950034YYxxxxxxxx}
 \issn{1362-3044}
\issnp{0950-0340} \jvol{xx} \jnum{xx} \jyear{2013} \jmonth{xx}

\markboth{Taylor \& Francis and I.T. Consultant}{Journal of Modern Optics}

\title{Interplay between radiation pressure force and scattered light intensity in the cooperative scattering by cold atoms}

\author{ T. Bienaim\'e$^{a}$, R. Bachelard$^{b}$, J. Chab\'e$^{a}$, M. T. Rouabah$^{a, c}$, L. Bellando$^{a}$, Ph. W. Courteille$^{b}$, N. Piovella$^{d}$, R. Kaiser$^{a}$
$^{\ast}$\vspace{6pt}\thanks{$^\ast$Corresponding author. Email:
robin.kaiser@inln.cnrs.fr \vspace{6pt}}
\\\vspace{6pt}
$^{a}${\em{Universit\'e de Nice Sophia Antipolis, CNRS, Institut Non-Lin\'eaire de Nice, UMR 7335, F-06560 Valbonne, France}}\\
$^{b}${\em{Instituto de F\'isica de S\~ao Carlos, Universidade de S\~ao Paulo, 13560-970 S\~ao Carlos, SP, Brazil}}
$^{c}${\em{Laboratoire de Physique Math\'ematique et Physique Subatomique, Universit\'e Constantine 1, Constantine 25000, Algeria}}\\
$^{d}${\em{Dipartimento di Fisica, Universit\`a Degli Studi di Milano, Via Celoria 16, I-20133 Milano, Italy}}
\\\vspace{6pt}\received{\today} }

\maketitle

\begin{abstract}
The interplay between the superradiant emission of a cloud of cold two-level atoms and the radiation pressure force is discussed.
Using a microscopic model of coupled atomic dipoles driven by an external laser, the radiation field and the average radiation pressure force are derived. A relation between the far-field scattered intensity and the force is derived, using the optical theorem. Finally, the scaling of the sample scattering cross section with the parameters of the system is studied.
\bigskip
\end{abstract}

\section{Introduction}

Cooperative effects occur when the behavior of a many body system is determined by their collective interactions with each other and thus manifest themselves in a large variety of physical systems. In this paper, we focus on the specific case of a collection of atoms illuminated by a laser. In this situation, the electro-magnetic field mediates resonant dipole-dipole interactions between the atoms, leading to a cooperative response of the system, which quantitatively differs from the single atom response. Such effects are imprinted on physical observables that can be experimentally measured such as e.g. the emission diagram or the radiation pressure force acting on the cloud.

When a single atom is illuminated by a laser, the scattering
process results in a force proportional to the number of scattered photons. Indeed, as an atom absorbs a photon from the laser of
wave vector $\mathbf{k}_0$, it acquires a momentum $\hbar
\mathbf{k}_0$, but the average momentum change during the emission
process is zero.

For a collection of atoms, the picture changes drastically as it was first noticed in a pioneering work by Dicke~\cite{Dicke} where he showed enhanced spontaneous emission decay rates in small and large samples due to constructive interferences of collective emission. In the situation of an incident laser scattering on a cloud of atoms, the atoms cooperate to scatter the light leading to a directional emission. This phenomenon is due to the synchronization of the atomic dipoles with the laser. The collective effects becomes even stronger as the atomic medium
becomes optically dense and the radiation of the atoms starts to alter significantly the wave propagation. Among the other collective
effects that arise, one can mention the collective Lamb shift~\cite{FHM,Keaveney2012}, Mie resonances~\cite{Bachelard2012a}, subradiance~\cite{Bienaime2012}, the refractive index of a dilute Bose gas \cite{Morice95} as well as a reduction of the radiation pressure force~\cite{Courteille2010,Bienaime2010}.

Since the radiated light results from the interference of the
waves emitted by each dipole, the simple relation between emitted
photon and atomic recoil is lost. For example, a striking feature
of cooperativity is the modification of the atomic recoil due to
the presence of the neighboring
atoms~\cite{Campbell2005,Bachelard2012b}, an effect that cannot be
deduced from single-atom physics.

We here discuss the particular relation between the directional superradiant emission, and the reduction of the radiation pressure force.
The atomic cloud is described as a microscopic ensemble of coupled atomic dipoles, and both the radiated field and the force are expressed
as a function of these dipoles. The optical theorem is derived in this framework, and is shown to lead to a direct relation between intensity
scattered and radiation pressure force for the cloud center-of-mass.

\section{Cooperative scattering model}

The atomic cloud is described as a system of two-level ($g$ and $e$) atoms, with resonant frequency
$\omega_a$ and position $\mathbf{r}_j$, that are driven by an uniform laser
beam with electric field amplitude $E_0$, frequency $\omega_0$ and
wave vector $\mathbf{k}_0=(\omega_0/c)\mathbf{\hat e}_z$. The laser-atom interaction is
described by the following Hamiltonian:
\begin{eqnarray}\label{H}
H&=&\frac{\hbar\Omega_0}{2}\sum_{j=1}^N\left[\hat\sigma_j
e^{i(\Delta_0 t- \mathbf{k}_0\cdot \mathbf{r}_j)}+\textrm{h.c.}\right]\nonumber\\
&+& \hbar\sum_{j=1}^N\sum_{\mathbf{k}}g_k\left(\hat\sigma_j
e^{-i\omega_a t} +\hat\sigma_j^\dagger e^{i\omega_a t}\right)
\left[\hat a_{\mathbf{k}}^\dagger e^{i(\omega_k t- \mathbf{k}\cdot
\mathbf{r}_j)}+\hat a_{\mathbf{k}} e^{-i(\omega_k t-
\mathbf{k}\cdot \mathbf{r}_j)}\right]
\end{eqnarray}
where $\Omega_0=d E_0/\hbar$ is the Rabi frequency of the incident
laser field and $\Delta_0=\omega_0-\omega_a$ is the detuning
between the laser and the atomic transition. In Eq. (\ref{H})
$\hat\sigma_j=|g_j\rangle\langle e_j|$ is the lowering operator
for $j-$atom, $\hat a_{\mathbf{k}}$ is the photon annihilation
operator and $g_k=(d^2\omega_k/2\hbar\epsilon_0 V)^{1/2}$ is the
single-photon Rabi frequency, where $d$ is the electric-dipole
transition matrix element and $V$ is the photon mode volume. The
special case where a single photon (mode
$\mathbf{k}$) can be assumed to be present in the system, was
extensively investigated in Refs.~\cite{FHM,Scully2006,Svi08}, and later extended to include a low-intensity laser in Ref.~\cite{Courteille2010,Bachelard2011,Bienaime2011}. The
system atoms+photons is then described by a state of the
form~\cite{Svi10}:
\begin{eqnarray}\label{state}
    |\Psi\rangle&=&\alpha(t)|g_1\dots g_N\rangle |0\rangle_{\mathbf{k}}+e^{-i\Delta_0 t}\sum_{j=1}^N
    \beta_j(t)|g_1\ldots e_j\ldots g_N\rangle|0\rangle_{\mathbf{k}}+ \sum_{\mathbf{k}}\gamma_{\mathbf{k}}(t)|g_1\dots g_N\rangle
    |1\rangle_{\mathbf{k}}\nonumber\\
    &+&\sum_{\mathbf{k}}\sum_{m,n=1}^N
    \epsilon_{m<n,\mathbf{k}}(t)|g_1\ldots e_m\ldots e_n\ldots
    g_N\rangle|1\rangle_{\mathbf{k}},
\end{eqnarray}
The first term in Eq. \eqref{state} corresponds to the initial
ground state without photons, the second term is the sum over the
states where a single atom has been excited by the classical
field. The third term corresponds to the atoms that returned to
the ground state having emitted a photon in the mode $\mathbf{k}$,
whereas the last one corresponds to the presence of two excited
atoms and one virtual photon with `negative' energy. It is due to
the counter-rotating terms in the Hamiltonian (\ref{H}) and
disappears when the rotating wave approximation is made.
 In the linear regime $\alpha\approx 1$ and in the Markov
 approximation, valid if the decay time is larger than the
 photon time-of-flight through the atomic cloud, the scattering problem reduces to the following differential
 equation~\cite{Scully09,Bachelard2011,Bienaime2011}
 \begin{equation}\label{eqbetaj}
    \dot\beta_j=\left(i\Delta_0-\frac{\Gamma}{2}\right)\beta_j- i\frac{\Omega_0}{2}e^{i \mathbf{k}_0\cdot
    \mathbf{r}_j}-\frac{\Gamma}{2}\sum_{m\neq j}
    \frac{\exp(ik_0|\mathbf{r}_j-\mathbf{r}_m|)}{ik_0|\mathbf{r}_j-\mathbf{r}_m|}\beta_m
\end{equation}
with initial condition $\beta_j(0)=0$, for $j=1,\dots,N$. Here,
$\Gamma=V g_k^2 k_0^2/\pi c=d^2k_0^3/2\pi\epsilon_0\hbar$ is the
single-atom {\it spontaneous} decay rate. The kernel in the last
term of Eq. (\ref{eqbetaj}) has a real component,
$-(\Gamma/2)\sum_{m\neq j}[\sin(x_{jm})/x_{jm}]$ (where
$x_{jm}=k_0|\mathbf{r}_j-\mathbf{r}_m|$), describing the {\it
collective} atomic decay, and an imaginary component,
$i(\Gamma/2)\sum_{m\neq j}[\cos(x_{jm})/x_{jm}]$, describing the
collective Lamb shift~\cite{Scully09,Scully10,Ralfie}.  Notice that while
Eq. (\ref{eqbetaj}) is here deduced from a quantum mechanical
model, it can also be obtained classically, treating the two-level
atoms as weakly excited classical harmonic
oscillators~\cite{Svi10,Prasad}.

\section{Radiated field}

The radiation field operator $\hat a_{\mathbf{k}}$ evolves
according to the following Heisenberg equation
\begin{equation}\label{aH}
    \frac{d\hat
    a_{\mathbf{k}}}{dt}=\frac{1}{i\hbar}[\hat a_{\mathbf{k}},\hat
    H]=-ig_k e^{i(\omega_k-\omega_a) t}\sum_{m=1}^N \hat\sigma_m e^{-i\mathbf{k}\cdot
    \mathbf{r}_m},
\end{equation}
where the fast oscillating term proportional to
$\exp[i(\omega_k+\omega_a)t]$ has been neglected. The scattered
field is obtained by performing the sum over all the modes, considering only the positive-frequency part of the
electric field operator
\begin{equation}\label{Es:a}
    \hat E_s(\mathbf{r},t)=\sum_{\mathbf{k}}{\cal E}_{k}\hat
    a_{\mathbf{k}}(t)
    e^{i\mathbf{k}\cdot \mathbf{r}-i\omega_k t}
\end{equation}
where ${\cal E}_k=(\hbar\omega_k/2\epsilon_0 V)^{1/2}$.
Integrating Eq. (\ref{aH}) with respect to time, with $a_{\mathbf{k}}(0)=0$, inserting it
in Eq. (\ref{Es:a}), and assuming the usual Markov approximation, one obtains~\cite{Bienaime2011}
\begin{equation}\label{Es:3}
    \hat E_s(\mathbf{r},t)\approx -\frac{dk_0^3}{4\pi\epsilon_0}e^{-i\omega_at}\sum_{m=1}^N
    \frac{e^{ik_0 |\mathbf{r}-\mathbf{r}_m|}}{k_0|\mathbf{r}-\mathbf{r}_m|}
    \hat\sigma_m(t).
\end{equation}
When applied on the state~\eqref{state}, neglecting virtual
transitions, it yields $\hat
E_s|\Psi\rangle=E_s\exp(-i\omega_0t)|g_1\dots g_N\rangle$, with
\begin{equation}\label{Es}
    E_s(\mathbf{r},t)=
    -\frac{\hbar \Gamma}{2d}\sum_{m=1}^N
    \beta_m(t)  \frac{e^{ik_0 |\mathbf{r}-\mathbf{r}_m|}}{k_0|\mathbf{r}-\mathbf{r}_m|}
\end{equation}
Hence, the radiated field appears as a sum of spherical waves radiated by the atomic dipoles. In the far-field limit,
one has $k_0|\mathbf{r}-\mathbf{r}_m|\approx k_0r-\mathbf{k}\cdot\mathbf{r}_m$, with
$\mathbf{k}=k_0(\mathbf{r}/r)$, so the field~\eqref{Es} radiated in a direction $\mathbf{k}$ reads
\begin{equation}\label{Es:far}
    E_s^{\mathrm{(far)}}(\mathbf{k},t)\approx
    -\frac{\hbar \Gamma}{2d}\frac{e^{ik_0r}}{k_0 r}\sum_{m=1}^N
    \beta_m(t)e^{-i\mathbf{k}\cdot\mathbf{r}_m}.
\end{equation}

The scattered intensity in a direction $\mathbf{k}$ is then derived as
\begin{eqnarray}\label{Es:far2}
    I_s(\mathbf{k})&=& \frac{\epsilon_0 c\hbar^2 \Gamma^2}{2(dk_0 r)^2}\left|\sum_{m=1}^N \beta_m(t)e^{-i\mathbf{k}\cdot\mathbf{r}_m}\right|^2
    \\ &=&\frac{\epsilon_0 c\hbar^2 \Gamma^2}{2(dk_0 r)^2}\left(\sum_{m=1}^N |\beta_m|^2+\sum_{j \neq m}^N \beta_{j}\beta_{m}^*
    e^{-i\mathbf{k}\cdot(\mathbf{r}_j-\mathbf{r}_m)}\right).
\end{eqnarray}

Integrating this intensity over all directions leads to the total scattered power
\begin{equation}\label{eq:P}
    P_{r}=\frac{d^2k_0^4c}{2\pi\epsilon_0} \left(\sum_{m=1}^N |\beta_m|^2+\sum_{m \neq j}^N \beta_{j}\beta_{m}^* \frac{\sin(k_0|\mathbf{r}_j-\mathbf{r}_m|)}{k_0|\mathbf{r}_j-\mathbf{r}_m|}\right),
\end{equation}
where we have used the equality
\begin{equation}
\int \mbox{d}\hat{\mathbf{k}} e^{ik_0
\hat{\mathbf{k}}\cdot\mathbf{d}}=4\pi\frac{\sin(k_0|d|)}{k_0|d|}.
\end{equation}
In Eq. \eqref{eq:P}, the first term corresponds to the {\it incoherent} sum of the single atom radiated power.
The second term is an interference term; in the limit of a cloud small compared to the wavelength, the dipole moments have
the same phase and this latter term is responsible for a superradiant build-up of the radiated power $\propto N^2$~(see, e.g., Ref.~\cite{Dicke}).

\section{Radiation pressure force}

As for the radiation force operator acting on the $j${th} atom, it is
derived from Eq. (\ref{H}) as
\begin{equation}
\hat{\mathbf{F}}_j=-\nabla_{\mathbf{r}_j}\hat H=\hat{\mathbf{F}}_{aj}+\hat{\mathbf{F}}_{ej}.
\end{equation}
A first contribution associated to the absorption of photons of the pump appears~\cite{Courteille2010,Bachelard2011}:
\begin{equation}
    \hat{\mathbf{F}}_{aj}= i\hbar \mathbf{k}_0\frac{\Omega_0}{2}
    \left\{\hat\sigma_{j} e^{i(\Delta_0 t-\mathbf{k}_0\cdot \mathbf{r}_j)}-
    \textrm{h.c.}\right\},\label{Force-abs}
\end{equation}
whereas the second contribution comes from the emission of the photons in any
direction $\mathbf{k}$:
\begin{equation}
   \hat{\mathbf{F}}_{ej}= i\hbar\sum_{\mathbf{k}} \mathbf{k}g_{k}
    \left\{\hat a_{\mathbf{k}}^\dagger \hat\sigma_{j}
    e^{i(\omega_k-\omega_a)t-i\mathbf{k}\cdot\mathbf{r}_j}-
    \hat\sigma_{j}^\dagger \hat a_{\mathbf{k}}
    e^{-i(\omega_k-\omega_a)t+i\mathbf{k}\cdot\mathbf{r}_j}\right\}.
    \label{Force-emi}
\end{equation}
In Eq. \eqref{Force-emi}, the
counter-rotating terms proportional to $\exp[\pm
i(\omega_k+\omega_a)t]$ were neglected.

As we are interested in comparing the radiation pressure force to
the single-atom case, we define the average radiation force
$\hat{\mathbf{F}}=(1/N)\sum_j\hat{ \mathbf{F}}_j=(F_{tot}/N)
\mathbf{\hat e}_z$ that measures acceleration of the cloud
center-of-mass given by $\mathbf{a}_{CM}=\hat{\mathbf{F}}/m$, with
$m$ the single-atom mass. Note that this average force is $N$
times smaller than the total force $F_{tot}$ acting on the whole
cloud of atoms. Since we consider clouds with rotational symmetry
around the laser axis, this force is in the same direction as the
incident field wave vector $\mathbf{k}_0=k_0 \mathbf{\hat e}_z$.
This average force is measured by time-of flight techniques in
cold atomic clouds released, for instance, from magneto-optical
traps (MOTs) and has recently revealed cooperative effects in the
scattering by extended atomic
samples~\cite{Bienaime2010,Bender2010}. Like the scattered
radiation, this force is an observable that contains signatures of
the cooperative scattering by the
atoms~\cite{Courteille2010,Bienaime2010}. The average absorption
force along the $z$-axis, resulting from the recoil received upon
absorption of a photon from the incident laser, reads
\begin{eqnarray}
 \hat F_a &=&
 \frac{i}{2N}\hbar k_0\Omega_0\sum_{j=1}^N\left[\hat\sigma_j e^{i\Delta_0 t-i \mathbf{k}_0\cdot
 \mathbf{r}_j}-\textrm{h.c.}\right].
 \label{Fa}
\end{eqnarray}
Similarly, the average emission force writes
$\hat{\mathbf{F}}_e=(1/N)\sum_j\hat {\mathbf{F}}_{ej}$. Inserting
the expression for $\hat a_{\mathbf{k}}$ from Eq. \eqref{aH} into
Eq. \eqref{Force-emi}, and approximating the discrete sum over the
modes $\mathbf{k}$ by an integral, it is possible to obtain, as it
was done for the radiation field operator $\hat E_S$ of
Eq. (\ref{Es}), the following expression for the average emission
force along the $z$-axis~\cite{Courteille2010}:
\begin{eqnarray}
   \hat F_e &=& -\frac{\hbar k_0\Gamma}{8\pi N}\int_0^{2\pi}\mbox{d}\phi\int_0^\pi
  \mbox{d}\theta\sin\theta\cos\theta\sum_{j,m=1}^N\left[
    e^{-i\mathbf{k}\cdot(\mathbf{r}_j-\mathbf{r}_m)}
    \hat\sigma_{m}^\dagger\hat\sigma_{j}+\textrm{h.c.}\right].
  \label{Fez}
\end{eqnarray}
Neglecting virtual photon contributions, the expectation values of
the absorption and emission forces for state \eqref{state} are
\begin{eqnarray}
 \langle\hat F_a\rangle &=& -\frac{\hbar k_0\Omega_0}{N} \sum_{j=1}^N\textrm{Im}\left[\beta_j e^{-i \mathbf{k}_0\cdot
 \mathbf{r}_j}\right]
 \label{Fazj} \\
    \langle\hat F_{e}\rangle&=& -\frac{\hbar k_0\Gamma}{4\pi N}\int_0^{2\pi}\mbox{d}\phi\int_0^\pi
  \mbox{d}\theta\sin\theta\cos\theta\sum_{j,m=1}^N\left[\beta_{j}\beta_{m}^*
    e^{-i\mathbf{k}\cdot(\mathbf{r}_j-\mathbf{r}_m)}\right]\nonumber\\
    &=& -\frac{\hbar k_0\Gamma}{N}
    \sum_{j,m=1}^N
    \frac{(z_j-z_m)}{|\mathbf{r}_j-\mathbf{r}_m|}j_1(k_0|\mathbf{r}_j-\mathbf{r}_m|)\mathrm{Im}\left(\beta_j\beta_m^*\right),
    \label{Fezj}
\end{eqnarray}
where  we used the identity
\begin{equation}
\int_0^{2\pi}\mbox{d}\phi\int_0^\pi
  \mbox{d}\theta\sin\theta\cos\theta
  e^{-i\mathbf{k}\cdot(\mathbf{r}-\mathbf{r}')}=4\pi
  i\frac{z-z'}{|\mathbf{r}-\mathbf{r}'|}j_1(k_0|\mathbf{r}-\mathbf{r}'|).\label{id:j1}
\end{equation}
$j_1(z)$ here refers the first order spherical Bessel function.
Note that the decomposition into absorption \eqref{Fazj} and
emission \eqref{Fezj} forces is fully compatible with classical
expressions of the optical force~\cite{Piovella2013}, where the
force arises as the product between the atomic dipole and the {\it
total} field~\cite{Gordon1980} (i.e. including the radiation from
the other atoms).

\section{Optical Theorem}\label{OT}

\begin{figure}[t]
\centering{\includegraphics[height=8cm]{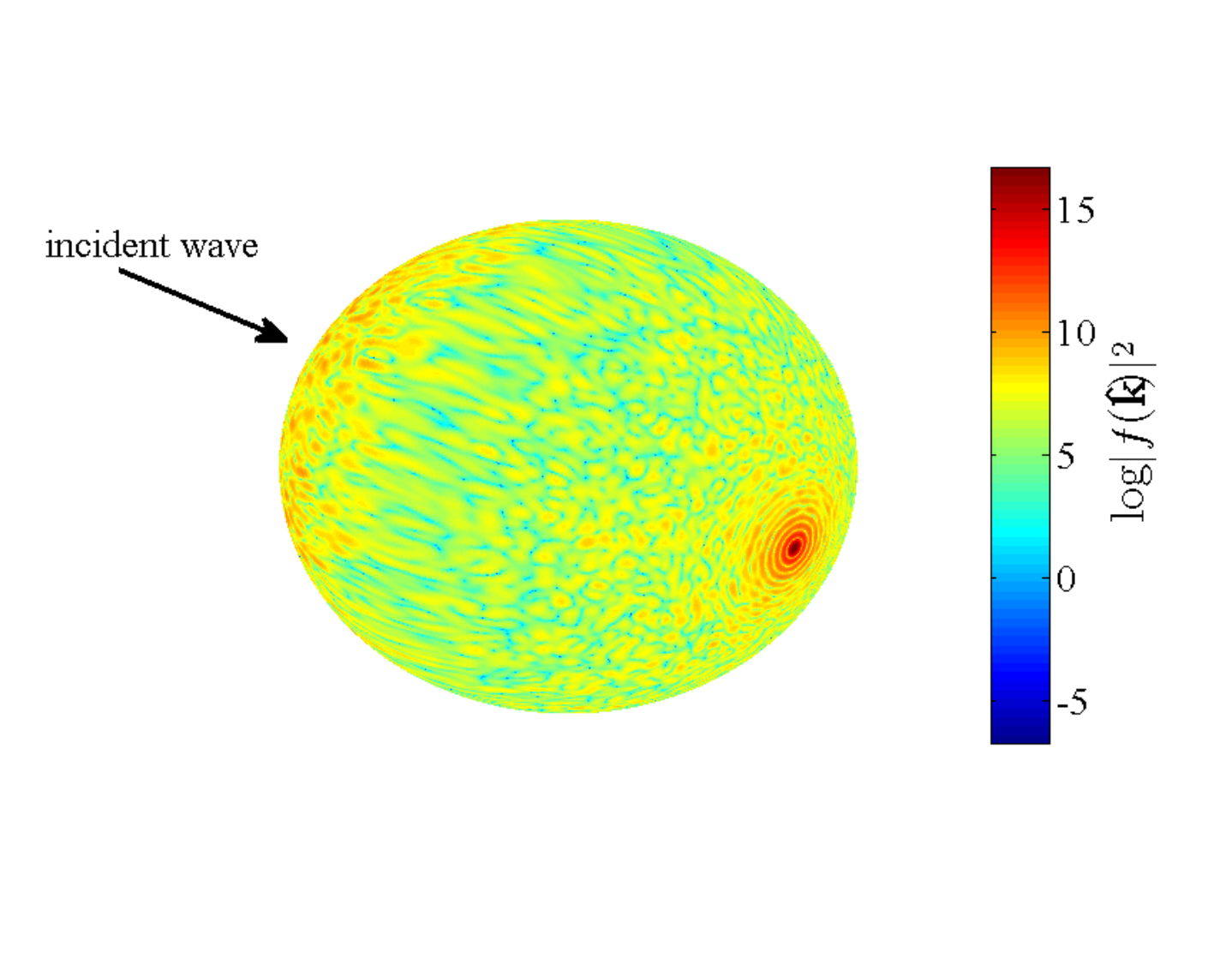}}
\caption{ Scattering amplitude $|f(\mathbf{\hat{k}})|^2$ as given by
Eq. (\ref{f}) for a cylindrical cloud of thickness $30/k_0$ and radius $90/k_0$, shone by a plane wave. The direction of the incoming wave is indicated by an arrow. The number of scatterers is $N=20000$, the detuning $\Delta_0=0$. The color-coded
intensity is represented in log-scale. One can clearly see in red
the strong forward emission of the sample, reminiscent of Mie
scattering by large clouds compared to the wavelength. In the
other directions, the scattered field is speckle-like due to the
randomly positioned two-level scatterers, and describes the
spontaneous emission by the cloud. Performing configuration
averages would smooth out these fluctuations, except in the
backward direction where, in the multiple scattering regime, the
well known coherent backscattering cone is
recovered~\cite{Lagendijk85,Maret85}. Finally, the emission in the transverse dimension is reduced due to the quasi-one-dimensional geometry.} \label{Emission_Diagram}
\end{figure}
 
\begin{figure}[t]
\centering{\includegraphics[height=6cm]{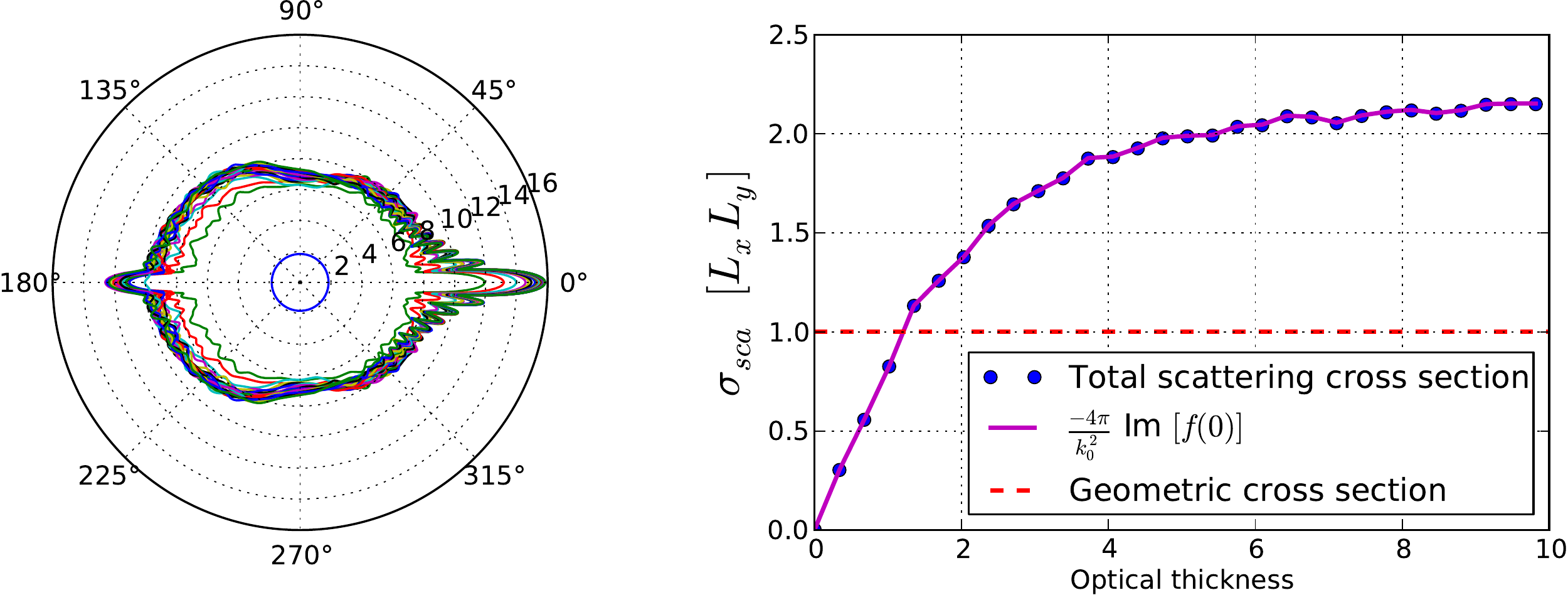}}
\caption{Illustration of the optical theorem. Left: the scattered
intensity integrated along $\phi$, i.e., $g (\theta) = \int_0^{2
\pi} d \phi \, |f(\theta,\phi)|^2$, is shown for resonant light
$\Delta_0=0$ and a slab geometry with a uniform density
distribution. The number of atomic scatterers is varied between
$1$ and $5000$ (from inside to outside curves). The transverse
size of the slab is $L_{x,y}=80/k_0$ and the longitudinal size is
varied such that $L_z = (20/k_0)N/5000$. This procedure allows us
to vary the optical thickness $b_0 = 4 \pi N / (k_0^2 L_x L_y)$
between $3.10^{-3}$ and $10$ while maintaining the atomic density
constant.  We would like to insist on the fact that the optical thickness is computed for the scattering of a scalar field which leads to an unusual resonant cross section for light $\sigma_0 = \lambda^2/\pi$ (different from the well-know resonant cross section $\sigma_0 = 3 \lambda^2 / (2 \pi)$ for vectorial light). The incident field is coming from the left and the
intensity is plotted in log-scale. In addition to the forward
Mie-like lobe, a lobe is also observed in the backward direction
which we attribute to light reflection due to the sharp variation
of optical index when the light hits the slab. Right: the blue
circles represents the total scattering cross section obtained by
integrating the emission diagram over $\theta$ and $\phi$, i.e.,
$\mathcal \sigma_{sca} = 1/k_0^2 \times \int_0^\pi d \theta \,
\sin(\theta) g(\theta) $. In our microscopic model, there is no
absorption so that $\sigma_{abs} = 0$, leading to $\sigma_{ext} =
\sigma_{sca}$. The optical theorem Eq. (\ref{sext}) can thus be
written as $\sigma_{sca} = - (4 \pi / k_0^2) \textrm{Im}[f(0)]$,
which is plotted in magenta. The good agreement between the two
curves illustrates the validity of the optical theorem.}
\label{Optical_Theorem}
\end{figure}

Let us now discuss the formulation of the optical theorem in the framework of collective scattering.
To that purpose, we consider an infinite slab illuminated by a plane wave. In the far-field limit, the field in a direction $\mathbf{\hat{k}}$ is
\begin{equation}\label{Etot}
    E(\mathbf{r})=\left[\frac{E_0}{2}e^{i k_0 z}+
    E_s^{\mathrm{(far)}}(r,\mathbf{\hat{k}})\right]e^{-i\omega_0
    t}=\frac{E_0}{2}\left[e^{ik_0z}-\frac{e^{ik_0r}}{k_0r}f(\mathbf{\hat{k}})\right]e^{-i\omega_0 t}
\end{equation}
where the scattering amplitude for the scattered field $f$ is given by
\begin{equation}\label{f}
    f(\mathbf{\hat{k}})=\frac{\Gamma}{\Omega_0}\sum_j\beta_j e^{-ik_0\mathbf{\hat{k}}\cdot
    \mathbf{r}_j}.
\end{equation}
As a consequence, the scattered intensity at a large distance $r$ from the
cloud is
\begin{equation}\label{Is}
    I_s=I_0\frac{|f(\mathbf{\hat{k}})|^2}{k_0^2r^2},
\end{equation}
while the total scattering cross section is obtained by integrating over all the solid angle
\begin{equation}\label{Cs}
    \sigma_{sca}=\frac{1}{k_0^2}\int\mbox{d}\mathbf{\hat{k}}
    |f(\mathbf{\hat{k}})|^2.
\end{equation}
To simulate numerically the slab illuminated by a plane wave, we consider a cylinder of transverse size large compared to its thickness and to the wavelength, with a random homogeneous distribution of atoms. Figure~\ref{Emission_Diagram} shows the emission diagram of the scattered field for resonant excitation and a cylindrical cloud of atoms.
The energy conservation imposes that
\begin{equation}\label{energy}
    \sigma_{ext}=\sigma_{sca}+\sigma_{abs}
\end{equation}
where $\sigma_{ext}$ and $\sigma_{abs}$ are the cross sections for
extinction and absorption, respectively. The extinction cross
section is then obtained from the optical theorem. In the forward
direction the total field is
\begin{equation}\label{Efor}
    E_{fwd}(\theta=0)=\frac{E_0}{2}\left[e^{ik_0z}-\frac{e^{ik_0r}}{k_0r}f(0)\right]e^{-i\omega_0
    t}.
\end{equation}
In the slab configuration, the cloud radiates mainly in a narrow forward cone - the angle of the cone of emission is given by the inverse of the cloud transverse size. Hence, observing the field in a plane far from the atoms and within the forward cone of emission, the
radius expands as $r\approx z+(x^2+y^2)/2z$, and one obtains
\begin{equation}\label{Efor2}
    E_{fwd}(\mathbf{r})\approx\frac{E_0}{2}\left[1-\frac{f(0)}{k_0z}e^{ik_0(x^2+y^2)/2z}\right]e^{i(k_0z-\omega_0 t)}.
\end{equation}
So the intensity reads
\begin{equation}\label{Efor3}
    |E_{fwd}(\mathbf{r})|^2\approx
    \frac{|E_0|^2}{4}\left\{1-\frac{2}{k_0z}\textrm{Re}\left[f(0)e^{ik(x^2+y^2)/2z}\right]\right\},
\end{equation}
since we have neglected the quadratic term $|E_s|^2$. The measured intensity is the incident intensity minus the
extinction intensity. In Eq. \eqref{Efor3}, the integration over $x,\ y$ yields a factor
$2i\pi z/k_0$, and one gets
\begin{equation}\label{sext}
    \sigma_{ext}=-\frac{4\pi}{k_0^2}\textrm{Im}[f(0)].
\end{equation}
Hence, from Eq. \eqref{Cs} one obtains the relation
\begin{equation}\label{sext2}
    -\textrm{Im}[f(0)]=\frac{1}{4\pi}\int \mbox{d}\mathbf{\hat{k}}
    |f(\mathbf{\hat{k}})|^2+\frac{k_0^2}{4\pi}\sigma_{abs}
\end{equation}
In our microscopic description of the light-atom interaction there
is no absorption, so that $\sigma_{abs}=0$. An illustration of the
validity of the optical theorem is given in Figure
\ref{Optical_Theorem} for resonant light scattering by a slab
containing two-level scatterers with a uniform density
distribution. From Eqs. \eqref{f} and \eqref{sext}, and introducing the wavevector $\mathbf{k}=k_0\mathbf{\hat{k}}(\theta,\phi)$, we obtain the
relation
\begin{equation}\label{Csbeta}
    -\frac{\Omega_0}{\Gamma}\sum_j\textrm{Im}\left[\beta_j e^{-i\mathbf{k}_0\cdot \mathbf{r}_j}\right]=
    \frac{1}{4\pi}\int_0^{2\pi}\mbox{d}\phi\int_0^\pi \mbox{d}\theta\sin\theta
    \sum_{j,m}\left[\beta_{j}\beta_{m}^*
    e^{-ik_0\mathbf{\hat{k}}\cdot(\mathbf{r}_j-\mathbf{r}_m)}\right]
\end{equation}
Consequently, using Eqs. \eqref{Fazj} and \eqref{Fezj}, the average force along the $z$-axis reads:
\begin{eqnarray}\label{force}
 F_z &=& \frac{\hbar k_0\Gamma}{4\pi N}\int_0^{2\pi}\mbox{d}\phi\int_0^\pi
  \mbox{d}\theta\sin\theta(1-\cos\theta)\sum_{j,m=1}^N\left[\beta_{j}\beta_{m}^*
    e^{-i\mathbf{k}\cdot(\mathbf{r}_j-\mathbf{r}_m)}\right].
\end{eqnarray}
We observe from Eq. \eqref{force} that the average radiation pressure force is not merely
proportional to the excitation probability, i.e.
$\sum_j|\beta_j|^2$, but it is the result of an interference
between the different atomic dipoles $\beta_j$. For this reason a
measurement of the force captures the coherence properties of
the scattering process as well as the detection of the light
intensity. To make this point more explicit, using Eq. \eqref{Es:far2}, it is possible to write
the force as
\begin{eqnarray}\label{force2}
 F_z &=&
 \frac{r^2}{Nc}\int_0^{2\pi}\mbox{d}\phi\int_0^\pi
  \mbox{d}\theta\sin\theta(1-\cos\theta)I_s(\theta,\phi),
\end{eqnarray}
where  the scattered far-field intensity is
$I_s(\theta,\phi)=2c\epsilon_0|E_s(\theta,\phi)|^2$. This highlights
the fact that the radiation pressure force, that pushes the atoms along the
direction of the incident beam, is proportional to the net
radiation flux of the scattered intensity.

In the case of an isotropic emission (e.g., single-atom case, or
cloud much smaller than the wavelength), the scattered intensity
$I_s$ is independent on the angle and we get $F_z=(4\pi
r^2/(Nc))I_s$: the direct proportionality between scattered
power and radiation pressure force is recovered. The cooperative effect of light scattering in such small samples is then encoded in the total scattered intensity $I_s$. In the case
of superradiant scattering for larger samples, a pronounced emission into the forward
direction decreases the radiation force, as observed for example
in Ref.~\cite{Bienaime2010}.

\section{Scaling of the scattering cross section}

\begin{figure}[t]
\centering{\includegraphics[height=6cm]{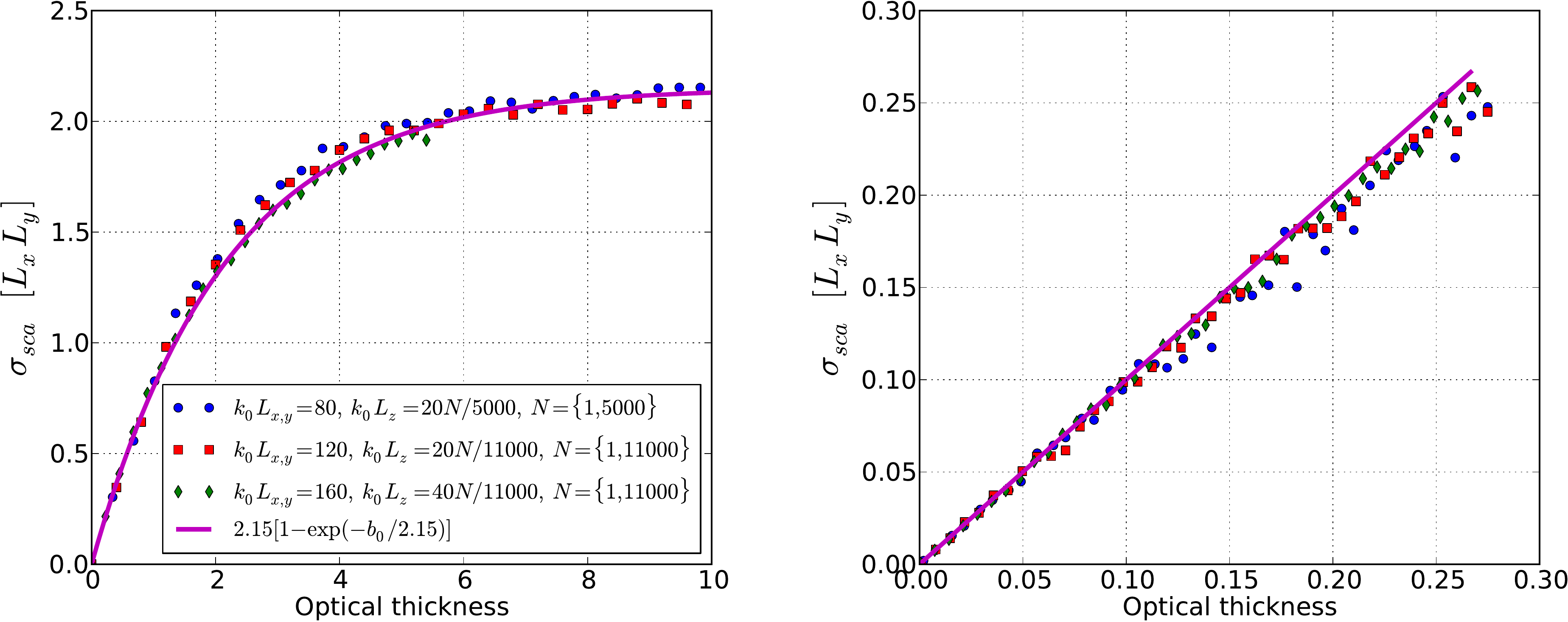}}
\caption{Scaling of the scattering cross section. Left plot:
following the same procedure as the one described in figure
\ref{Optical_Theorem}, we compute the scattering cross sections
for different slab geometries. The results are shown in scatter
plot with different colors. The parameters of the simulations are
reported in the legend of the figure. By fitting the data,
constraining the slope in the limit $b_0\rightarrow 0$ (right
plot), we obtain a scattering cross section that scales with the
optical thickness $b_0$ of the slab according to Eq. (\ref{fit})
(magenta full line).} \label{Cross_Section_Scaling}
\end{figure}

In this section we are interested in understanding how the
scattering cross section scales with the parameters of the system.
We consider the case of a slab with uniform density distribution.
The slab contains $N$ atoms and its size along the $x$, $y$, $z$
axes is denoted by $L_x$, $L_y$, $L_z$ respectively. The numerical
simulations presented in figure \ref{Cross_Section_Scaling} show
how the scattering cross section depends on the optical thickness
of the cloud $b_0 = 4 \pi N / (k_0^2 L_x L_y)$. For dilute clouds
of atoms we find:

\begin{equation}
\sigma_{sca} = 2.15\times L_x L_y \left[ 1 - \exp \left(
-\frac{b_0}{2.15} \right) \right].\label{fit}
\end{equation}

When the slab is optically thick, i.e. $b_0 \gg 1$, we observe
that the cross section appears to approach $2\times L_x L_y$. This
factor of two corresponds to the well-known ``extinction paradox"
\cite{Hulst81,Bohren83} for which the extinction cross section is
twice as large as the one predicted by geometrical optics due to
the diffraction contribution. The residual deviations from the
factor of 2 between the scattering and geometrical cross sections
might be associated to a still moderate size of our sample \cite{Chomaz11}, or to dipole blockade effects~\cite{Ott2013,Bienaime2013}. For
spherical dielectric spheres, $\sigma_{ext}$ shows an oscillatory
behavior around $2 \sigma_{geo}$ ($\sigma_{geo}=L_xL_y$ for our square geometry),
which is damped for increasing sizes of the sphere \cite{Kargl90,
Berg11}. When $b_0 \ll 1$ the
scattering cross section can be written as $\sigma_{sca} = (L_x L_y)b_0 =  N \sigma_0$, where
$\sigma_0= \lambda^2/ \pi$ is the resonant scattering cross section for a single atom in the scalar wave description (it differs from the well know cross section for vectorial light $\sigma_0 = 3 \lambda^2 / (2 \pi)$). In this limit, the interpretation is clear: at low optical thickness the cooperative effects are negligible and the scattering of light is given by the response of $N$ independent atoms. We refer the reader to \cite{Sokolov13} for a study of the areal scaling of the light scattering by varying the size of a dense, cold atomic cloud.

Before concluding, we would like to underline the importance of the role of diffraction. Since we are using a microscopic description of the system, diffraction effects for the scattered field are already included in our model. However, free propagation of the incident field needs to be added for a fully consistent description. In this respect, the incident plane wave considered so far in the paper is a peculiar case. We will focus on these aspects in forthcoming studies to precisely understand the role of diffraction. This will naturally lead us to compare our coherent microscopic model of coupled dipoles to stochastic incoherent models commonly used to describe photon propagation in random media. Understanding coherent light propagation in disordered resonant scatterers is of prime importance for both the atomic physics and the waves in complex media communities.

\section{Conclusion}

We here discussed the superradiant emission of a cloud of cold
atoms, when the interference of the waves radiated by the atomic
dipoles builds up a coherent emission. Despite the fact that the
simple relation between absorbed photons and radiation pressure
force existing in the single-atom case was lost, the optical
theorem allowed to recover a simple relation between the total
scattered intensity and the displacement of the cloud
center-of-mass. The measure of the force of the center of mass of
the atomic cloud contains (partial) information on the scattered
intensity, even for large values of optical thickness of the
cloud. We have computed the total scattering cross section which
approaches a value close to twice the geometrical cross section of
the sample, in line with the well-know extinction paradox. Finally, understanding the role of diffraction paves the way for further studies to compare our coherent microscopic model to well established stochastic incoherent models describing photon propagation in random media.

\section{Acknowledgements}

We acknowledge financial support from IRSES project COSCALI and from USP/COFECUB (projet Uc Ph 123/11). R. B. and Ph. W. C. acknowledge support from the Funda\c{c}\~ao de Amparo à Pesquisa do Estado de S\~ao Paulo (FAPESP). M. T. R. is supported by an Averro\`es exchange program.

\end{document}